\documentclass[pra,showpacs,showkeys,amsfonts,amsmath]{revtex4}
\usepackage{graphicx}

\begin{document}

\title{Generalizing {T}sirelson's bound on {B}ell inequalities using a min-max principle}

\author{Stefan Filipp}
\email{sfilipp@ati.ac.at}
\affiliation{Atominstitut der {\"{O}}sterreichischen Universit{\"{a}}ten,
Stadionallee 2, A-1020 Vienna, Austria}
\author{Karl Svozil}
\email{svozil@tuwien.ac.at}
\homepage{http://tph.tuwien.ac.at/~svozil}
\affiliation{Institut f\"ur Theoretische Physik, University of Technology Vienna,
Wiedner Hauptstra\ss e 8-10/136, A-1040 Vienna, Austria}

\begin{abstract}
Bounds on the norm of quantum operators associated with classical Bell-type inequalities
can be derived from their maximal eigenvalues.
This quantitative method enables detailed
predictions of the maximal violations of Bell-type inequalities.
\end{abstract}


\maketitle

The violations of Bell-type inequalities represent
a cornerstone of our present understanding of quantum probability theory
\cite{peres}.
Thereby, the usual procedure is as follows:
First, the (in)equalities bounding the classical probabilities and expectations are
derived systematically; e.g., by enumerating all conceivable classical possibilities
and their associated two-valued measures.
These form the extreme points which span
the classical correlation polytopes
\cite{cirelson:80,cirelson,froissart-81,pitowsky-86,pitowsky,pitowsky-89a,Pit-91,Pit-94,2000-poly,collins-gisin-2003,sliwa-2003};
the faces of which are expressed by Bell-type inequalities
which characterize the bounds of the classical probabilities and expectations;
in Boole's term \cite{Boole,Boole-62}, the ``conditions of possible experience.''
(Generating functions are another method to find bounds on classical expectations \cite{werner-wolf-2001,schachner-2003}.)
The Bell-type inequalities contain sums of (joint) probabilities and expectations.
In a second step, the classical probabilities and expectations in
the Bell-type inequalities are substituted by quantum probabilities and expectations.
The resulting operators violate the classical bounds.
Until recently, little was known about the fine structure of the violations.
{T}sirelson (also written Cirel'son) published an absolute bound for the violation of a particular Bell-type inequality,
the Clauser-Horne-Shimony-Holt (CHSH) inequality \cite{cirelson:80,cirelson:87,cirelson,khalfin-97}.
Cabello has investigated a violation of the CHSH inequality beyond the quantum mechanical
bound by applying selection schemes to particles in a GHZ-state
\cite{cabello-02a,cabello-02b}.
Recently, detailed numerical  \cite{filipp-svo-04-qpoly}
and analytical studies \cite{cabello-2003a} stimulated
experiments \cite{bovino-2003} to test the quantum bounds of certain Bell-type inequalities.

In what follows,
a general method to compute quantum bounds on Bell-type inequalities
will be reviewed systematically.
It makes use of the {\em min-max principle} for self-adjoint transformations
(Ref.~\cite{halmos-vs}, Sec.~90 and  Ref.~\cite{reed-sim4}, Sec.~75)
stating that the operator norm is bounded by the minimal and maximal eigenvalues.
These ideas are not entirely new and have been mentioned previously
\cite{werner-wolf-2001,filipp-svo-04-qpoly,cabello-2003a},
yet to our knowledge no systematic investigation has been undertaken yet.
It should also be kept in mind that this method {\it a priori}
cannot produce quantum polytopes \cite{pit:range-2001,filipp-svo-04-qpoly},
but the quantum correspondents of classical polytopes.
Indeed, as we demonstrate explicitly, the resulting geometric forms are not convex.
This, however,
does not diminish the relevance of these quantum predictions
to experiments testing the quantum violations
of classical Bell-type inequalities.

As a starting point note that
since $(A+B)^\dagger =A^\dagger +B^\dagger = (A+B)$ for arbitrary self-adjoint transformations $A,B$,
the sum of self-adjoint transformations is again self-adjoint.
That is, all self-adjoint transformations entering the quantum correspondent of any Bell-type inequality
is again a self-adjoint transformation.
The sum does not preserve eigenvectors and eigenvalues;
i.e., $A+B$ can have different eigenvectors and eigenvalues than $A$ and $B$ taken separately
(i.e., $A$ and $B$ need not necessarily commute).
The norm of the self-adjoint transformation resulting from summing the quantum counterparts
of all the classical terms contributing to a particular Bell inequality obeys the min-max principle.
Thus determining the maximal violations of classical Bell inequalities amounts to
solving an eigenvalue problem.
The associated eigenstates are the multi-partite states which yield a maximum violation
of the classical bounds under the given experimental (parameter) setup
\footnote{
Nondegenerate eigenstates are always representable by one-dimensional subspaces
and thus are pure, the exception being the possibility of a mixing between
degenerate eigenstates \cite{braunstein92}.}.

Let us demonstrate the method by considering two particles propagating in inverse directions;
their polarization or spin being
measured along two and more ($m$)
distinct directions per particle
perpendicular to their propagation directions.
For these configurations, we shall enumerate analytical quantum bounds corresponding to the
Clauser-Horne (CH) inequality, as well as of more general inequalities
 for $m > 2$ \cite{2000-poly,collins-gisin-2003,sliwa-2003}.

For $m=2$, the CH inequalities restrict classical probabilities by
$-1 \leq p_{13} + p_{14} + p_{23} - p_{24}- p_{1} -p_{3} \leq 0$, as well as permutations thereof.
Here, $p_{1}$ and $p_{3}$ stand for the probabilities that the first particle is measured along
the first direction and the second particle is measured along the third direction.
$p_{ij}$ stands for the joint probability to find the first particle along the direction $i$
and the second particle along the direction $j$.

In order to evaluate the quantum counterpart of the CH inequalities,
the classical probabilities have to be substituted by the quantum ones;
i.e.,
\begin{equation}
\begin{array}{lll}
p_1 &\rightarrow& q_1 (\theta ) =
{\frac{1}{2}}\left[{\mathbb I}_2 + {\bf \sigma}( \theta )\right] \otimes  {\mathbb I}_2,
\\
p_3 &\rightarrow& q_3 (\theta ) =
{\mathbb I}_2 \otimes {\frac{1}{2}}\left[{\mathbb I}_2 + {\bf \sigma}( \theta )\right],
\\
p_{ij}&\rightarrow& q_{ij} (\theta ,\theta ') =
{\frac{1}{2}}\left[{\mathbb I}_2 + {\bf \sigma}( \theta )\right]
\otimes
{\frac{1}{2}}\left[{\mathbb I}_2 + {\bf \sigma}( \theta ')\right],
\end{array}
\label{2004-qbounds-e2}
\end{equation}
with
$
{\bf \sigma}( \theta )=
\begin{pmatrix} \cos \theta & \sin \theta  \\
  \sin\theta & -\cos \theta
  \end{pmatrix}
$,
where $\theta $ is the relative measurement angle in the $x$--$z$-plane, and the two particles propagate along the $y$-axis.
The quantum transformation associated with the CH inequality enumerated above is thus given by
\begin{equation}
\begin{array}{lcl}
O_{22}(\alpha,\beta,\gamma,\delta)
&=&  q_{13}(\alpha,\gamma) +
q_{14}(\alpha,\delta) + q_{23}(\beta,\gamma) - q_{24}(\beta,\delta)- q_{1}(\alpha) - q_{3}(\gamma)\\
&=&
{\frac{1}{2}}\left[{\mathbb I}_2 + {\bf \sigma}( \alpha )\right]
\otimes
{\frac{1}{2}}\left[{\mathbb I}_2 + {\bf \sigma}( \gamma )\right] +
{\frac{1}{2}}\left[{\mathbb I}_2 + {\bf \sigma}( \alpha )\right]
\otimes
{\frac{1}{2}}\left[{\mathbb I}_2 + {\bf \sigma}( \delta )\right] \\
&& +
{\frac{1}{2}}\left[{\mathbb I}_2 + {\bf \sigma}( \beta )\right]
\otimes
{\frac{1}{2}}\left[{\mathbb I}_2 + {\bf \sigma}( \gamma )\right]
- {\frac{1}{2}}\left[{\mathbb I}_2 + {\bf \sigma}( \beta)\right]
\otimes
{\frac{1}{2}}\left[{\mathbb I}_2 + {\bf \sigma}( \delta)\right]\\
&& -
{\frac{1}{2}}\left[{\mathbb I}_2 + {\bf \sigma}( \alpha )\right] \otimes  {\mathbb I}_2
- {\mathbb I}_2 \otimes {\frac{1}{2}}\left[{\mathbb I}_2 + {\bf \sigma}( \gamma )\right],
\end{array}
\label{2004-qbounds-e4}
\end{equation}
where $\alpha$, $\beta$, $\gamma$, $\delta$ denote the measurement angles
lying in the $x$--$z$-plane: $\alpha$ and $\beta$ for one particle, $\gamma$ and
$\delta$ for the other one.
The eigenvalues of the self-adjoint transformation in
(\ref{2004-qbounds-e4}) are
\begin{equation}
  \label{eq:2004-qbounds-evo22}
  \lambda_{1,2,3,4}(\alpha,\beta,\gamma,\delta )
=
\frac{1}{2}\big(\pm\sqrt{1\pm\sin(\alpha -\beta )\sin(\gamma -\delta )}-1\big)
\end{equation}
yielding the maximum bound
$\|O_{22} \|= \max_{i=1,2,3,4} \lambda_i$.
Note that for the particular choice of parameters
$\alpha=0,\beta=2\theta,\gamma=\theta,\delta=3\theta$ adopted in
\cite{cabello-2003a,filipp-svo-04-qpoly}, one obtains $|O_{22}|=
\frac{1}{2}\left\{\left[ \left(3-\cos 4\theta \right)/2\right]^{1/2}
  -1\right\}$.

In the Bell-basis
$\{|\phi^+ \rangle,|\psi^+ \rangle,|\psi^- \rangle,|\phi^- \rangle\}$ with
$|\psi^\pm \rangle = 1/\sqrt{2}(|01 \rangle \pm |10 \rangle)$ and
$|\phi^\pm \rangle = 1/\sqrt{2}(|00 \rangle \pm |11 \rangle)$,
the eigenvectors corresponding to the maximal violating
eigenstates are
\begin{equation}
\begin{array}{ccl}
  |\nu_\pm \rangle&=&\big(F^\pm(\alpha,\beta,-\gamma ,-\delta  ) |\psi^+ \rangle +
  |\phi^- \rangle\big)\big(1+F^\pm(\alpha,\beta ,-\gamma,-\delta )^2\big)^{-\frac{1}{2}},\\
  |\mu_\pm \rangle&=&\big(F^\pm(\alpha,\beta,\gamma,\delta )|\phi^+ \rangle +
  |\psi^- \rangle\big)\big(1+F^\pm(\alpha ,\beta,\gamma,\delta )^2\big)^{-\frac{1}{2}},
\end{array}
\label{eq:2004-qbounds-evo22es}
\end{equation}
with
$$
F^\pm(\alpha ,\beta ,\gamma,\delta )=\pm 2\sqrt{1-\sin(\alpha -\beta )\sin(\gamma -\delta )}\;\;
 \frac{\cos(\alpha -\delta )-\cos(\alpha -\gamma )-\cos(\beta -\gamma )-\cos(\beta -\delta )}
{\sin(\alpha -\gamma )+\sin(\beta -\gamma )-\sin(\alpha -\delta )
  +\sin(\beta -\delta )}.
$$
The states (\ref{eq:2004-qbounds-evo22es})
are maximally entangled, corroborating the approach of
Cabello \cite{cabello-2003a} to utilize a set of
maximally entangled states to
reconstruct the quantum bound for the setting of the relative angles
$\alpha=0$, $\beta=2\theta$, $\gamma=\theta$ and $\delta=3\theta$ \footnote{
Equivalent results hold for the Clauser-Horne-Shimony-Holt (CHSH) inequality
 \cite{mermin-1995,cereceda-2001}.}.
>From the
particular form of the eigenstates, we conclude that the maximal
violating eigenstates of the $O_{22}$  operator are maximally
entangled for general measurement angles lying in the $x$--$z$-plane.

Generalizations for $m$ measurements per particle are
straightforward;
for example, the extension to \emph{three} measurement operators for each particle
yields only one additional nonequivalent (with respect to symmetries)
inequality \cite{collins-gisin-2003,sliwa-2003}
$I_{33}=p_{14} + p_{15} + p_{16} + p_{24} + p_{25} - p_{26} + p_{34} - p_{35}
- p_{1} - 2 p_{4} - p_{5} \leq 0$ among the 684 inequalities \cite{2000-poly} representing the
 faces of the associated classical correlation polytope.
The associated operator for symmetric
measurement directions is given by
\begin{widetext}
\begin{equation}
\begin{array}{lll}
&O_{33}(0,\theta,2\theta,0,\theta,2\theta)= q_{14}(0,0) + q_{15}(0,\theta) + q_{16}(0,2\theta) + q_{24}(\theta,0) +
q_{25}(\theta,\theta) - q_{26}(\theta,2\theta) +\\
&\qquad  + q_{34}(2\theta,\theta)- q_{35}(2\theta,\theta)-q_{1}(0) - 2 q_{4}(0) - q_{5}(\theta) \\
&\qquad \qquad =\frac{1}{4}\left(
\begin{smallmatrix}
-4\sin^2\theta & 0 & 0 & 0\\
0 & -5-2\cos\theta - 3\cos 2\theta + 2\cos 3\theta &
4\cos^2\frac{\theta}{2} & 2\sin\theta + 3 \sin 2\theta - 2 \sin
3\theta\\
0 & 4\cos^2\frac{\theta}{2} & -2(3+\cos 2\theta) & - 2\sin\theta \\
0 &  2\sin\theta + 3 \sin 2\theta - 2 \sin 3\theta & - 2\sin\theta &
2\sin^2\frac{\theta}{2}\cos^2\frac{\theta}{2}(4\cos\theta -3)
\end{smallmatrix}\right),
\end{array}
\label{2004-qbounds-e5}
\end{equation}
\end{widetext}
again in the Bell basis and for quantum expressions similar to the ones enumerated in Eq.~(\ref{2004-qbounds-e2})
\footnote{Note that  $q_{1}$ refers to the state of the first particle,
whereas $q_{4}$ and $q_{5}$ refer to the states of the second
particle.}.

\begin{figure}[htbp]
  \centering
  \includegraphics[width=90mm]{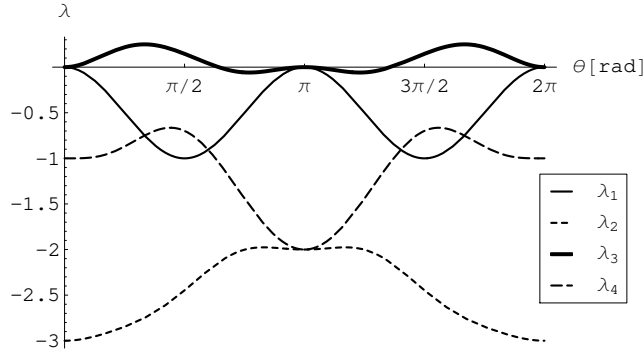}
  \caption{Eigenvalues of $O_{33}$ in dependence of the relative angle $\theta$.}
  \label{fig:2004-qbounds-f1}
\end{figure}

In this basis, the operator $O_{33}(0,\theta,2\theta,0,\theta,2\theta)$
splits into a direct sum of a one-dimensional part $-\sin^2\theta $
and a three-dimensional part $o$, respectively. Using the Cardano method
(see Ref.~\cite{cocolicchio00}), one can
solve the characteristic equation of the three dimensional submatrix
$o$ in the lower right corner of $O_{33}$
\begin{equation}
  \lambda^3 + b(\theta) \lambda^2 + c(\theta) \lambda + d(\theta) = 0,
\label{2004-qbounds-characteristic}
\end{equation}
with the coefficients $b = -\text{Tr}\, o,\ c = 1/2\Big(\text{Tr}\,^2 o -
\text{Tr}\, o^2 \Big),\ d = -\det o$. (For convenience we omit here the
dependence on $\theta$.) The (real) eigenvalues can then be written as \cite{cocolicchio00}
\begin{eqnarray}
\lambda_2 = -2 \sqrt{|u|}\cos\frac{\xi}{3}-\frac{b}{3}\nonumber\\
\lambda_{3,4} = \sqrt{|u(x)|}\Big[\cos\frac{\xi}{3} \pm
\sin\frac{\xi}{3}\Big]-\frac{b}{3},
\label{2004-qbounds-o33ev}
\end{eqnarray}
with $u=1/9(3 c-b^2)$ and $\cos\xi = \frac{1}{54}\big(9 b c -2 b^3 - 27 d\big)/\big(u\sqrt{|u|}\big)$.
In Fig. \ref{fig:2004-qbounds-f1},
the eigenvalues $\lambda_2, \lambda_3, \lambda_4$,
together with the eigenvalue $\lambda_1 = -\sin^2\theta$ from the
one-dimensional part of $O_{33}$, are plotted  as functions of the parameter $\theta$ .
The maximum violation of $1/4$ is obtained for $\theta=\pi/3$ with the associated
eigenvector
\begin{equation}
  |\Psi_{\rm max } \rangle=\frac{\sqrt{3}}{2}|\phi^- \rangle+\frac{1}{2}|\psi^+ \rangle.
\label{2004-qbounds-pmax33}
\end{equation}

\begin{figure}[htbp]
  \centering
  \includegraphics[width=90mm]{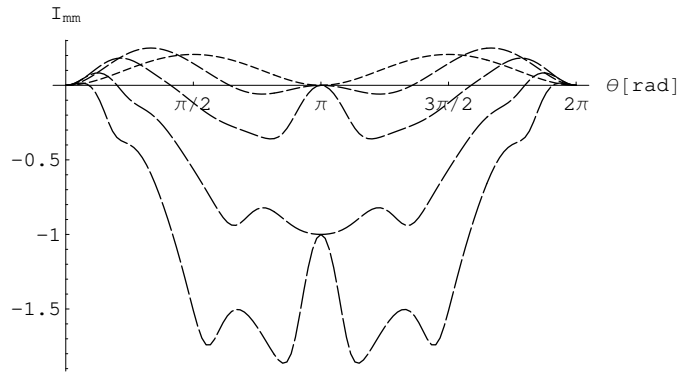}
  \caption{Maximum violation of the operator $O_{mm}$ for
    $m=2,\ldots,6$ for a symmetric measurement setup; longer dashes indicate larger $m$.}
  \label{fig:2004-qbounds-f2}
\end{figure}

As indicated in Ref.~\cite{collins-gisin-2003}, this scheme can be extended to $m$
measurements on each particle by considering inequalities $I_{mm}
\leq 0$ and
corresponding operators $O_{mm}$ of the form
\begin{eqnarray}
  I_{mm}&=& \sum_{j=1}^{m}\sum_{i=1}^{m-j+1}P({A_i B_j})-\sum_{i=1}^{m-1}
  P({A_{i+1}B_{m-i+1}}) \nonumber\\
  &&-\sum_{i=1}^{m}(m-i)P(B_{i}) - P(A_1) \leq 0,
\end{eqnarray}
where $P(A_i B_j)$ denotes the joint probability of obtaining the value one of the
projection operators $A_i$ and $B_j$ operators on the left and on the
right hand side, and $P(A_i),\ P(B_j)$ the marginal probabilities on
one side, respectively.
For a choice of measurement directions
$\{0,\theta,2\theta,\ldots,m\theta\}$ on both sides,
the maximizing eigenvalues
are plotted in Fig. \ref{fig:2004-qbounds-f2}.
The matrices belonging to the operators $O_{mm}$ ($m \leq
6$)
are of the same form
as is $O_{33}$, i.~e. they
split up into a direct sum of two matrices in the
Bell-basis; the maximal eigenvalues can therefore be  calculated
explicitly using Eqs. (\ref{2004-qbounds-characteristic}) and (\ref{2004-qbounds-o33ev}).


For experimental realizations of the $O_{33}$ case and special parameter configurations,
the {\it ansatz} of Cabello \cite{cabello-2003a} and Bovino {\it et al.} \cite{bovino-2003}
can be generalized to arbitrary {\em local}
unitary transformations $U_{2\times 2} \in SU(2) \otimes SU(2)$ applied to each one of the two particles
in some Bell-basis state $|\varphi \rangle$
separately; i.e.,
\begin{equation}
  U(\omega_1,\theta_1,\phi_1)\otimes U(\omega_2,\theta_2,\phi_2) |\varphi \rangle .
\label{2004-qbounds-gencab}
\end{equation}
The single
qubit operators are taken as $U(\omega,\theta,\phi) =
e^{i\frac{\omega}{2} \vec{n}\cdot\vec{\sigma}} \in SU(2)$ with
$\omega$ as the rotation angle about the axis $\vec{n}=
(\sin\theta\cos\phi,\sin\theta\sin\phi,\cos\theta)^T$.
For example, the use of the Bell state
$|\psi^+ \rangle $ and
and the successive application of the local unitary operation $U(\omega_1,\theta_1,\phi_1)\otimes
U(\omega_2,\theta_2,\phi_2)$ with $\omega_1 =2\pi/3$,
$\theta_1=\phi_1=\pi/2$ and $\omega_2=\theta_2=\phi_2=0$ yields the  maximally violating eigenvector
$|\Psi_{\rm max }\rangle$ from
Eq. (\ref{2004-qbounds-pmax33}) which is also maximally entangled.

For the general $m>2$ case,  however,
it is not always possible to obtain all possible maximally violating states by starting from a Bell state:
for general measurement angles, the experimental realization additionally requires a two-qubit
transformation from
$SU(4)/(SU(2)\otimes SU(2))$, followed by a local
unitary operation $U_{2\times 2}$ in order to obtain all possible
states \cite{zhang-2003}.
As an example, consider the maximally violating but not maximally entangled state at $\theta=\pi/2$:
$|\Psi_{\pi/2} \rangle \approx 0.86
  |\psi^+ \rangle + 0.17|\psi^- \rangle + 0.47|\phi^- \rangle$ cannot be
  obtained from a Bell state, as
  entanglement is preserved under $SU(2) \otimes SU(2)$ operations.

Alternatively, multiport interferometry \cite{rzbb,zukowski-97,svozil-2004-analog}
offers a direct proof-of-principle
implementation:
By choosing the appropriate transmission coefficients and phases in a generalized
beam splitter setup, one can prepare any pure state
from an input state $|11 \rangle \equiv (0,0,0,1)$ corresponding to a photon
in a single input port. Take, for example, the maximal eigenstate of
the $O_{33}$ operator at $\theta=\pi/2$, $|\Psi_{\pi/2} \rangle \approx 0.86
  |\psi^+ \rangle + 0.17|\psi^- \rangle + 0.47|\phi^- \rangle \equiv
  (0.34,0.73,0.49,-0.34)$. The appropriate transmission parameters
  can be calculated via the identification \cite{rzbb}
  \begin{equation}
    \begin{pmatrix}0\\0\\0\\1\end{pmatrix}^T R(N)^{-1} =  \begin{pmatrix}0.34\\0.73\\0.49\\-0.34\end{pmatrix}^T =
    \begin{pmatrix}e^{-i\phi_1}\cos\omega_1 \\-e^{-i\phi_2}\cos\omega_2\sin\omega_1\\e^{-i\phi_3}\cos\omega_3\sin\omega_2\sin\omega_1\\-\sin\omega_3\sin\omega_2\sin\omega_1\end{pmatrix}^T
  \end{equation}
for $\omega_1 = 1.23$, $\omega_2=2.46$, $\omega_3=0.60$ and $\phi_1=\phi_2=\phi_3=0$,
where $R(N)$ is a
$SU(4)$ rotation serially composed by two-dimensional beamsplitter
matrices.

In summary, we have shown how to construct and experimentally test
the exact quantum bounds of
Bell-type inequalities
by solving the eigenvalue problem of the associated self-adjoint
transformation.
Several problems remain open.
Among them is
the exact derivation of the quantum correlation hull \cite{pit:range-2001,filipp-svo-04-qpoly},
in particular, whether the quantum hull is obtainable by extending the
classical Bell-type
inequalities in the way as presented above;
i.e., by substituting the quantum probabilities for the classical ones.
This is by no means trivial,
as the sections of the quantum hull need not necessarily be derivable by mere classical extensions.
A second open question is related to the geometric structures arising from quantum expectation values.
These need not necessarily be convex.
Again, the question of direct extensibility remains open for the hull of
quantum expectations from the classical ones.

This research has been supported by the Austrian Science Foundation (FWF),
Project Nr. F1513. S. F. acknowledges helpful conversations with
B. Hiesmayer and S. Scheel.


\begin{thebibliography}{32}
\expandafter\ifx\csname natexlab\endcsname\relax\def\natexlab#1{#1}\fi
\expandafter\ifx\csname bibnamefont\endcsname\relax
  \def\bibnamefont#1{#1}\fi
\expandafter\ifx\csname bibfnamefont\endcsname\relax
  \def\bibfnamefont#1{#1}\fi
\expandafter\ifx\csname citenamefont\endcsname\relax
  \def\citenamefont#1{#1}\fi
\expandafter\ifx\csname url\endcsname\relax
  \def\url#1{\texttt{#1}}\fi
\expandafter\ifx\csname urlprefix\endcsname\relax\def\urlprefix{URL }\fi
\providecommand{\bibinfo}[2]{#2}
\providecommand{\eprint}[2][]{\url{#2}}

\bibitem[{\citenamefont{Peres}(1993)}]{peres}
\bibinfo{author}{\bibfnamefont{A.}~\bibnamefont{Peres}},
  \emph{\bibinfo{title}{Quantum Theory: Concepts and Methods}}
  (\bibinfo{publisher}{Kluwer Academic Publishers},
  \bibinfo{address}{Dordrecht}, \bibinfo{year}{1993}).

\bibitem[{\citenamefont{{Cirel'son}}(1980)}]{cirelson:80}
\bibinfo{author}{\bibfnamefont{B.~S.} \bibnamefont{{Cirel'son (=Tsirel'son)}}},
  \bibinfo{journal}{Letters in Mathematical Physics}
  \textbf{\bibinfo{volume}{4}}, \bibinfo{pages}{93} (\bibinfo{year}{1980}).

\bibitem[{\citenamefont{{Tsirelson}}(1993)}]{cirelson}
\bibinfo{author}{\bibfnamefont{B.~S.} \bibnamefont{{Cirel'son (=Tsirel'son)}}},
  \bibinfo{journal}{Hadronic Journal Supplement} \textbf{\bibinfo{volume}{8}},
  \bibinfo{pages}{329} (\bibinfo{year}{1993}).

\bibitem[{\citenamefont{Froissart}(1981)}]{froissart-81}
\bibinfo{author}{\bibfnamefont{M.}~\bibnamefont{Froissart}},
  \bibinfo{journal}{Nuovo Cimento B} \textbf{\bibinfo{volume}{64}},
  \bibinfo{pages}{241} (\bibinfo{year}{1981}).

\bibitem[{\citenamefont{Pitowsky}(1986)}]{pitowsky-86}
\bibinfo{author}{\bibfnamefont{I.}~\bibnamefont{Pitowsky}},
  \bibinfo{journal}{J. Math. Phys.} \textbf{\bibinfo{volume}{27}},
  \bibinfo{pages}{1556} (\bibinfo{year}{1986}).

\bibitem[{\citenamefont{Pitowsky}(1989{\natexlab{a}})}]{pitowsky}
\bibinfo{author}{\bibfnamefont{I.}~\bibnamefont{Pitowsky}},
  \emph{\bibinfo{title}{Quantum Probability---Quantum Logic}}
  (\bibinfo{publisher}{Springer}, \bibinfo{address}{Berlin},
  \bibinfo{year}{1989}{\natexlab{a}}).

\bibitem[{\citenamefont{Pitowsky}(1989{\natexlab{b}})}]{pitowsky-89a}
\bibinfo{author}{\bibfnamefont{I.}~\bibnamefont{Pitowsky}}, in
  \emph{\bibinfo{booktitle}{{B}ell's Theorem, Quantum Theory and the
  Conceptions of the Universe}}, edited by
  \bibinfo{editor}{\bibfnamefont{M.}~\bibnamefont{Kafatos}}
  (\bibinfo{publisher}{Kluwer}, \bibinfo{address}{Dordrecht},
  \bibinfo{year}{1989}{\natexlab{b}}), pp. \bibinfo{pages}{37--49}.

\bibitem[{\citenamefont{Pitowsky}(1991)}]{Pit-91}
\bibinfo{author}{\bibfnamefont{I.}~\bibnamefont{Pitowsky}},
  \bibinfo{journal}{Mathematical Programming} \textbf{\bibinfo{volume}{50}},
  \bibinfo{pages}{395} (\bibinfo{year}{1991}).

\bibitem[{\citenamefont{Pitowsky}(1994)}]{Pit-94}
\bibinfo{author}{\bibfnamefont{I.}~\bibnamefont{Pitowsky}},
  \bibinfo{journal}{Brit. J. Phil. Sci.} \textbf{\bibinfo{volume}{45}},
  \bibinfo{pages}{95} (\bibinfo{year}{1994}).

\bibitem[{\citenamefont{Pitowsky and Svozil}(2001)}]{2000-poly}
\bibinfo{author}{\bibfnamefont{I.}~\bibnamefont{Pitowsky}} \bibnamefont{and}
  \bibinfo{author}{\bibfnamefont{K.}~\bibnamefont{Svozil}},
  \bibinfo{journal}{Physical Review A} \textbf{\bibinfo{volume}{64}},
  \bibinfo{pages}{014102} (\bibinfo{year}{2001}), \eprint{quant-ph/0011060},
  \urlprefix\url{http://dx.doi.org/10.1103/PhysRevA.64.014102}.

\bibitem[{\citenamefont{Colins and Gisin}(2003)}]{collins-gisin-2003}
\bibinfo{author}{\bibfnamefont{D.}~\bibnamefont{Colins}} \bibnamefont{and}
  \bibinfo{author}{\bibfnamefont{N.}~\bibnamefont{Gisin}}
  (\bibinfo{year}{2003}), \eprint{quant-ph/0306129}.

\bibitem[{\citenamefont{Sliwa}(2003)}]{sliwa-2003}
\bibinfo{author}{\bibfnamefont{C.}~\bibnamefont{Sliwa}},
  \bibinfo{journal}{Physics Letters A} \textbf{\bibinfo{volume}{317}},
  \bibinfo{pages}{165} (\bibinfo{year}{2003}), \eprint{quant-ph/0305190},
  \urlprefix\url{http://dx.doi.org/10.1016/S0375-9601(03)01115-0}.

\bibitem[{\citenamefont{Boole}(1958)}]{Boole}
\bibinfo{author}{\bibfnamefont{G.}~\bibnamefont{Boole}},
  \emph{\bibinfo{title}{An investigation of the laws of thought}}
  (\bibinfo{publisher}{Dover edition}, \bibinfo{address}{New York},
  \bibinfo{year}{1958}).

\bibitem[{\citenamefont{Boole}(1862)}]{Boole-62}
\bibinfo{author}{\bibfnamefont{G.}~\bibnamefont{Boole}},
  \bibinfo{journal}{Philosophical Transactions of the Royal Society of London}
  \textbf{\bibinfo{volume}{152}}, \bibinfo{pages}{225} (\bibinfo{year}{1862}).

\bibitem[{\citenamefont{Werner and Wolf}(2001)}]{werner-wolf-2001}
\bibinfo{author}{\bibfnamefont{R.~F.} \bibnamefont{Werner}} \bibnamefont{and}
  \bibinfo{author}{\bibfnamefont{M.~M.} \bibnamefont{Wolf}},
  \bibinfo{journal}{Phys. Rev. A} \textbf{\bibinfo{volume}{64}},
  \bibinfo{pages}{032112} (\bibinfo{year}{2001}), \eprint{quant-ph/0102024},
  \urlprefix\url{http://dx.doi.org/10.1103/PhysRevA.64.032112}.

\bibitem[{\citenamefont{Schachner}(2003)}]{schachner-2003}
\bibinfo{author}{\bibfnamefont{G.}~\bibnamefont{Schachner}}
  (\bibinfo{year}{2003}), \eprint{quant-ph/0312117}.

\bibitem[{\citenamefont{{Tsirel'son}}(1987)}]{cirelson:87}
\bibinfo{author}{\bibfnamefont{B.~S.} \bibnamefont{{Cirel'son (=Tsirel'son)}}},
  \bibinfo{journal}{Journal of Soviet Mathematics}
  \textbf{\bibinfo{volume}{36}}, \bibinfo{pages}{557} (\bibinfo{year}{1987}).

\bibitem[{\citenamefont{Khalfin and Tsirelson}(1992)}]{khalfin-97}
\bibinfo{author}{\bibfnamefont{L.~A.} \bibnamefont{Khalfin}} \bibnamefont{and}
  \bibinfo{author}{\bibfnamefont{B.~S.} \bibnamefont{Tsirelson}},
  \bibinfo{journal}{Foundations of Physics} \textbf{\bibinfo{volume}{22}},
  \bibinfo{pages}{879} (\bibinfo{year}{1992}).

\bibitem[{\citenamefont{Cabello}(2002{\natexlab{a}})}]{cabello-02a}
\bibinfo{author}{\bibfnamefont{A.}~\bibnamefont{Cabello}},
  \bibinfo{journal}{Physical Review Letters} \textbf{\bibinfo{volume}{88}},
  \bibinfo{pages}{060403} (\bibinfo{year}{2002}{\natexlab{a}}),
  \eprint{quant-ph/0108084}.

\bibitem[{\citenamefont{Cabello}(2002{\natexlab{b}})}]{cabello-02b}
\bibinfo{author}{\bibfnamefont{A.}~\bibnamefont{Cabello}},
  \bibinfo{journal}{Physical Review A} \textbf{\bibinfo{volume}{66}},
  \bibinfo{pages}{042114} (\bibinfo{year}{2002}{\natexlab{b}}),
  \eprint{quant-ph/0205183}.

\bibitem[{\citenamefont{Filipp and Svozil}(2004)}]{filipp-svo-04-qpoly}
\bibinfo{author}{\bibfnamefont{S.}~\bibnamefont{Filipp}} \bibnamefont{and}
  \bibinfo{author}{\bibfnamefont{K.}~\bibnamefont{Svozil}},
  \bibinfo{journal}{Physical Review A} \textbf{\bibinfo{volume}{69}},
  \bibinfo{pages}{032101} (\bibinfo{year}{2004}), \eprint{quant-ph/0306092},
  \urlprefix\url{http://dx.doi.org/10.1103/PhysRevA.69.032101}.

\bibitem[{\citenamefont{Cabello}(2004)}]{cabello-2003a}
\bibinfo{author}{\bibfnamefont{A.}~\bibnamefont{Cabello}},
  \bibinfo{journal}{Physical Review Letters} \textbf{\bibinfo{volume}{92}},
  \bibinfo{pages}{060403} (\bibinfo{year}{2004}), \eprint{quant-ph/0309172},
  \urlprefix\url{http://dx.doi.org/10.1103/PhysRevLett.92.060403}.

\bibitem[{\citenamefont{Bovino et~al.}(2004)\citenamefont{Bovino, Castagnoli,
   Degiovanni, Castelletto}}]{bovino-2003}
\bibinfo{author}{\bibfnamefont{F.~A.} \bibnamefont{Bovino}},
  \bibinfo{author}{\bibfnamefont{G.}~\bibnamefont{Castagnoli}},
  \bibinfo{author}{\bibfnamefont{I.~P.} \bibnamefont{Degiovanni}},
  \bibnamefont{and} \bibinfo{author}{\bibfnamefont{S.}~\bibnamefont{Castelletto}},
  \bibinfo{journal}{Physical Review Letters}
  \textbf{\bibinfo{volume}{92}}, \bibinfo{pages}{060404}
  (\bibinfo{year}{2004}), \eprint{quant-ph/0310042},
  \urlprefix\url{http://dx.doi.org/10.1103/PhysRevLett.92.060404}.

\bibitem[{\citenamefont{Halmos}(1974)}]{halmos-vs}
\bibinfo{author}{\bibfnamefont{P.~R.} \bibnamefont{Halmos}},
  \emph{\bibinfo{title}{Finite-dimensional vector spaces}}
  (\bibinfo{publisher}{Springer}, \bibinfo{address}{New York, Heidelberg,
  Berlin}, \bibinfo{year}{1974}).

\bibitem[{\citenamefont{Reed and Simon}(1978)}]{reed-sim4}
\bibinfo{author}{\bibfnamefont{M.}~\bibnamefont{Reed}} \bibnamefont{and}
  \bibinfo{author}{\bibfnamefont{B.}~\bibnamefont{Simon}},
  \emph{\bibinfo{title}{Methods of Modern Mathematical Physics IV: Analysis of
  Operators}} (\bibinfo{publisher}{Academic Press}, \bibinfo{address}{New
  York}, \bibinfo{year}{1978}).

\bibitem[{\citenamefont{Pitowsky}(2002)}]{pit:range-2001}
\bibinfo{author}{\bibfnamefont{I.}~\bibnamefont{Pitowsky}}, in
  \emph{\bibinfo{booktitle}{Quantum Theory: Reconsideration of Foundations,
  Proceedings of the 2001 V\"{a}xj\"{o} Conference}} (\bibinfo{publisher}{World
  Scientific}, \bibinfo{address}{Singapore}, \bibinfo{year}{2002}),
  \eprint{quant-ph/0112068}.

\bibitem[{\citenamefont{Cocolicchio and Viggiano}(2000)}]{cocolicchio00}
\bibinfo{author}{\bibfnamefont{D.}~\bibnamefont{Cocolicchio}} \bibnamefont{and}
  \bibinfo{author}{\bibfnamefont{D.}~\bibnamefont{Viggiano}},
  \bibinfo{journal}{J. Phys. A} \textbf{\bibinfo{volume}{33}},
  \bibinfo{pages}{5669} (\bibinfo{year}{2000}).

\bibitem[{\citenamefont{Zhang et~al.}(2003)\citenamefont{Zhang, Vala,
      Sastry, and Whaley}}]{zhang-2003}
\bibinfo{author}{\bibfnamefont{J.}~\bibnamefont{Zhang}},
\bibinfo{author}{\bibfnamefont{J.}~\bibnamefont{Vala}},
\bibinfo{author}{\bibfnamefont{S.}~\bibnamefont{Sastry}},
\bibnamefont{and}
\bibinfo{author}{\bibfnamefont{K.~B.}~\bibnamefont{Whaley}},
\bibinfo{journal}{Physical Review A} \textbf{\bibinfo{volume}{67}},
\bibinfo{pages}{042313} (\bibinfo{year}{2003}),
\urlprefix\url{http://dx.doi.org/10.1103/PhysRevA.67.042313}.

\bibitem[{\citenamefont{Reck et~al.}(1994)\citenamefont{Reck, Zeilinger,
  Bernstein, and Bertani}}]{rzbb}
\bibinfo{author}{\bibfnamefont{M.}~\bibnamefont{Reck}},
  \bibinfo{author}{\bibfnamefont{A.}~\bibnamefont{Zeilinger}},
  \bibinfo{author}{\bibfnamefont{H.~J.} \bibnamefont{Bernstein}},
  \bibnamefont{and} \bibinfo{author}{\bibfnamefont{P.}~\bibnamefont{Bertani}},
  \bibinfo{journal}{Physical Review Letters} \textbf{\bibinfo{volume}{73}},
  \bibinfo{pages}{58} (\bibinfo{year}{1994}),
  \urlprefix\url{http://dx.doi.org/10.1103/PhysRevLett.73.58}.

\bibitem[{\citenamefont{Zukowski et~al.}(1997)\citenamefont{Zukowski,
  Zeilinger, and Horne}}]{zukowski-97}
\bibinfo{author}{\bibfnamefont{M.}~\bibnamefont{Zukowski}},
  \bibinfo{author}{\bibfnamefont{A.}~\bibnamefont{Zeilinger}},
  \bibnamefont{and} \bibinfo{author}{\bibfnamefont{M.~A.} \bibnamefont{Horne}},
  \bibinfo{journal}{Physical Review A} \textbf{\bibinfo{volume}{55}},
  \bibinfo{pages}{2564} (\bibinfo{year}{1997}),
  \urlprefix\url{http://dx.doi.org/10.1103/PhysRevA.55.2564}.

\bibitem[{\citenamefont{Svozil}(2004)}]{svozil-2004-analog}
\bibinfo{author}{\bibfnamefont{K.}~\bibnamefont{Svozil}}
  (\bibinfo{year}{2004}), \eprint{quant-ph/0401113}.

\bibitem[{\citenamefont{Braunstein
      et~al.}(1992)\citenamefont{Braunstein, Mann, and
      Revzen}}]{braunstein92}
\bibinfo{author}{\bibfnamefont{S.~L.}~\bibnamefont{Braunstein}},
\bibinfo{author}{\bibfnamefont{A.}~\bibnamefont{Mann}}, \bibnamefont{and}
\bibinfo{author}{\bibfnamefont{M.}~\bibnamefont{Revzen}},
\bibinfo{journal}{Physical Review Letters}
\textbf{\bibinfo{volume}{68}},
\bibinfo{pages}{3259} (\bibinfo{year}{1992}).

\bibitem[{\citenamefont{Mermin}(1995)}]{mermin-1995}
\bibinfo{author}{\bibfnamefont{D.~N.} \bibnamefont{Mermin}},
  \bibinfo{journal}{Annals of the New York Academy of Sciences}
  \textbf{\bibinfo{volume}{755}}, \bibinfo{pages}{616} (\bibinfo{year}{1995}).

\bibitem[{\citenamefont{Cereceda}(2001)}]{cereceda-2001}
\bibinfo{author}{\bibfnamefont{J.~L.} \bibnamefont{Cereceda}},
  \bibinfo{journal}{Foundations of Physics Letters}
  \textbf{\bibinfo{volume}{14}}, \bibinfo{pages}{401} (\bibinfo{year}{2001}),
  \eprint{quant-ph/0101143}.

\end{thebibliography}

\end{document}